**Topological phonons in an inhomogeneously strained silicon-4: Large spin dependent thermoelectric response and thermal spin transfer torque due to topological electronic magnetism of phonons**

Ravindra G Bhardwaj[1‡], Anand Katailiha[1‡], Paul C. Lou[1], Ward P. Beyermann[2] and Sandeep Kumar[1, *]

[1] Department of Mechanical Engineering, University of California, Riverside, CA 92521, USA

[2] Department of Physics and Astronomy, University of California, Riverside, CA 92521, USA

[*] Corresponding author

Email: sandeep.suk191@gmail.com

Abstract

The superposition of flexoelectronic doping and topological phonons give rise to topological electronic magnetism of phonon in an inhomogeneously strained Si in the bilayer structure with metal. In case of ferromagnetic metal and Si bilayer structure, the flexoelectronic doping will also give rise to larger spin current, which will lead to large spin to charge conversion due to topological electronic magnetism of phonon. By applying a temperature difference to ferromagnetic metal/Si bilayer structure under an applied strain gradient, a large thermoelectric response can be generated. In this experimental study, we report a large spin dependent thermoelectric response at $Ni_{80}Fe_{20}$/Si bilayer structure. The spin dependent response is found to be an order of magnitude larger than that in Pt thin films and similar to topological insulators surface states in spite of negligible intrinsic spin-orbit coupling of Si. This large response is attributed to the flexoelectronic doping and topological electronic magnetism of phonons, which was uncovered using topological Nernst effect measurement. This alternative and novel approach of using inhomogeneous strain engineering to address both spin current density and spin to charge conversion can open a new window to the realization of spintronics and spin-caloritronics devices using metal and doped-semiconductor layered materials.

In metal-Si (highly doped) bilayer structure, interfacial flexoelectronic effect lead to charge carrier injection into Si layer, which will lead to flexoelectronic doping and asymmetric charge density in Si layer. The superposition of flexoelectronic doping and topological phonons in Si is reported to give rise to the topological electronic magnetism of phonons[1,2] demonstrated in the part 3 of this series and as shown in Figure 1(a). The temporal magnetic moment due to topological magnetism is expected to be the underlying cause of phonon skew scattering mediated spin-Hall effect (SHE)[3] and anomalous Hall effect (AHE)[2], reported previously. The spin-phonon coupling in Si was also demonstrated by the thermal conductivity measurements as a function of magnetic field[4]. As stated previously, the topological electronic magnetism arises due to interfacial flexoelectronic effect mediated charge carrier injection in Si.

In a ferromagnetic metal/Si bilayer structure, interfacial flexoelectronic effect[5] will cause flexoelectronic doping of Si layer in the absence of any magnetic field (randomized magnetic domains) similar to a normal metal. However, the flexoelectronic charge carrier injection will turn into a spin polarized charge current when an external magnetic field is applied. The resulting spin dependent electron-phonon interactions due to topological electronic magnetism of phonons will lead to spin to charge conversion from phonon skew scattering and transverse charge current. The research into spin dependent behavior has been restricted to designing better materials and interfaces for spin to charge conversion process. However, interfacial flexoelectronic effect gives an alternative for controlling the spin current density as well as spin to charge conversion using topological electronic magnetism of phonon for potential applications in spin-caloritronics, spintronics and waste heat recovery. We have already demonstrated large spin-Hall effect[3] using this approach. However, our approach can also be effective for spin dependent thermoelectric energy conversion. For example- in case of unstrained ferromagnetic metal/Si bilayer, a temperature gradient perpendicular to the interface will lead to small thermoelectric response as shown in Figure 1 (b). Whereas the spin current will be increased by order of magnitude due to flexoelectronic charge injection as shown in Figure 1 (c), which will enhance the spin dependent thermoelectric response significantly.

Previously, we uncovered the topological electronic magnetism of phonons using charge transport measurement methods. The thermoelectric response measurements can eliminate any charge transport effects in our previous measurement in part 1-3. This motivated us to undertake magneto-thermal transport measurements for thermoelectric energy conversion in flexoelectronic Si. In this study, we report a large spin dependent thermoelectric response in the case of Py/p-Si bilayer sample that was found to be an order of magnitude larger than that of Pt and similar to topological insulator surface states. This large spin dependent thermoelectric response is attributed to the flexoelectronic spin current and efficient spin to charge conversion due to the topological electronic magnetism of phonons in the Si layer. We also observe the interlayer coupling similar to Ruderman–Kittel–Kasuya–Yosida (RKKY), which is expected to arise due to flexoelectronic charge injection.

We designed and fabricated an experimental setup with integrated Pt heater using microelectromechanical systems (MEMS) approach[6,7] as shown in Figure 1 (d). The heater and the sample were electrically isolated using an intermediate MgO layer as shown in the high-resolution transmission electron microscope image in Figure 1 (f). The residual stresses, due to lattice mismatch and thermal expansion[8], in the MgO and Pt heater layer (Supplementary Figure S1) induced the required strain gradient in the sample, which led to delamination and cracking in some samples. The continuity of the thin film was verified using a HRTEM study (Supplementary Section-A) for the Py (25 nm)/Si (5 nm) sample, as shown in Figure 1 (e,f). Using HRTEM and AFM measurement, we deduced that the mean roughness for the p-Si samples is ~1.22 nm, as shown in Supplementary Figure S2. We fabricated[9,10](Supplementary Section A) four devices with the following sample structures: a Py (25 nm) control device, a Pt (3 nm)/Py (25 nm), a Py (25 nm)/p-Si (50 nm) and a Py (25 nm)/p-Si (5 nm). These devices will allow us to estimate the anomalous Nernst effect (ANE)[11,12] contribution from the Py layer and also evaluate the relative efficiency of the spin dependent thermoelectric response of the flexoelectronic Si. In the experimental measurement, the temperature gradient ($\Delta T_z$) was generated across the thickness of the thin film sample by passing an electric current (I) through the Pt heater, as shown in Figure 1 (d).

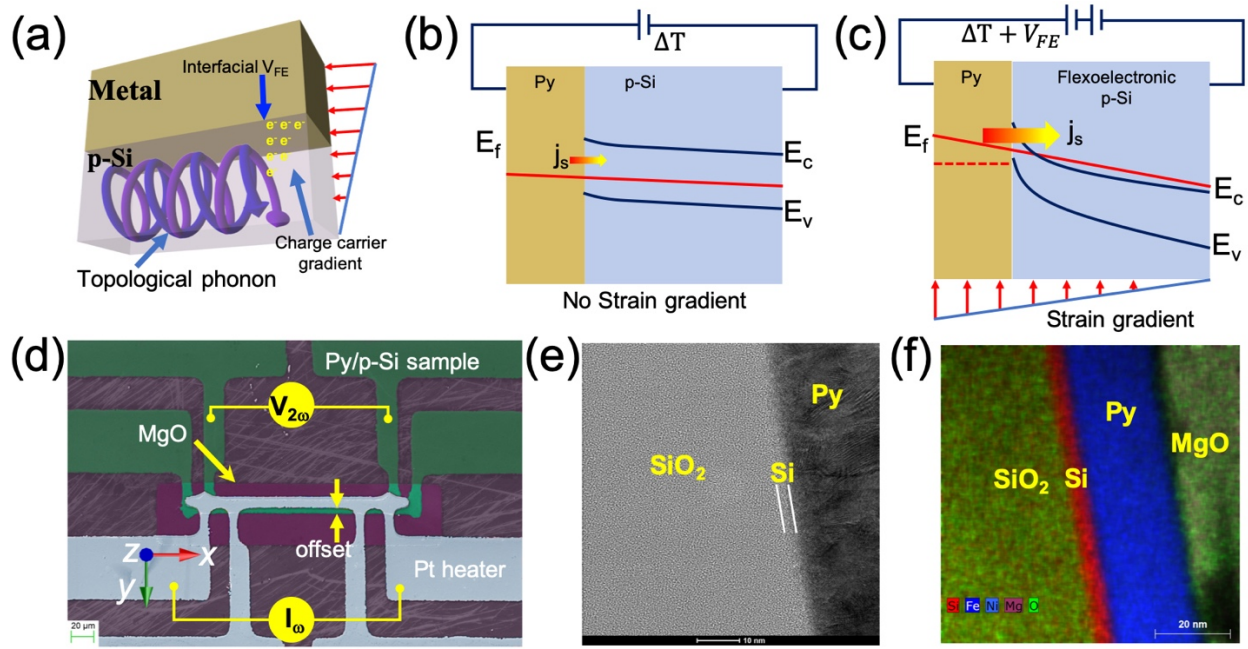

Figure 1. (a) schematic showing approximate band diagram of p-doped Si thin film due to flexoelectronic charge separation. (b) Schematic showing the band structure and charge carrier injection in the Py/Si bilayer thin film due to flexoelectronic effect. (c) A representative false color scanning electron micrograph showing the experimental device. (d) A high-resolution transmission electron micrograph showing the layered structure of the experimental specimen and (e) an energy dispersive X-ray spectroscopy elemental map showing the thin film layers and interfaces.

The transverse thermoelectric response measurements were undertaken at 30 mA/5 Hz of heating current and resulting $V_{2\omega}$ response (being quadratic in heating current) was acquired as a function of the magnetic field (1500 Oe to -1500 Oe) applied in the y-direction (normal to the temperature gradient), also known as in-plane magnetized (IM) configuration. The low frequency measurement eliminated any parasitic (capacitive and inductive) effects. The $V_{2\omega}$ responses were measured at 300 K as shown in Figure 2 (a). The Py and Si thin film resistivity were expected to be $5\times10^{-7}$ $\Omega m$ and $5\times10^{-5}$ $\Omega m$. We used the $3\omega$ method[13] to estimate the increase in heater temperature and finite element method simulation to estimate the temperature drop across Py layers as shown in Supplementary Section C and Supplementary Figure S3 and Supplementary Figure S4.

In the case of Py sample, we measured an ANE (easy axis) response of 15.1 µV where coefficient was estimated to be 0.1 µV/K as compared to the 0.045 µV/K[14] reported in literature. In the case of the Pt sample, the total response was ~33.1 µV and transverse spin dependent response due to spin-Seebeck effect (SSE) was estimated to be ~18 µV ; a behavior consistent with the reported ANE and SSE responses for the Pt/Py sample[15]. The spin-Seebeck coefficient in case of Pt/Py sample was 0.117 µV/K that was also larger than the reported value of ~74.3 nV/K[16]. The dimensional inaccuracies and self-induced spin dependent behavior at strained MgO and $SiO_2$ interfaces[17] could lead to the observed error. However, the error in coefficients indicated that the order of our temperature estimates was correct.

The transverse spin dependent thermal response in the case of the p-Si samples was measured to be ~85 µV and 164.45 µV for 50 nm and 5 nm Si, respectively, which was 4 times and 8.5 times larger than that of SSE response in Pt sample. This difference could not arise due to shunting effect since the resistivity of the Si was two order of magnitude larger than Py thin film. In addition, the shunting should have reduced the overall response as compared to Py sample. The estimated spin-dependent thermoelectric coefficients were $1.032 \pm 0.1$ µV/K and $0.458 \pm 0.05$ µV/K for 5 nm Si and 50 nm Si devices, respectively, (details in Supplementary Section C). The spin dependent thermoelectric response in 5 nm p-Si sample was larger than 50 nm p-Si sample, which was attributed to the larger strain gradient and the increased flexoelectronic doping in 5 nm p-Si sample.

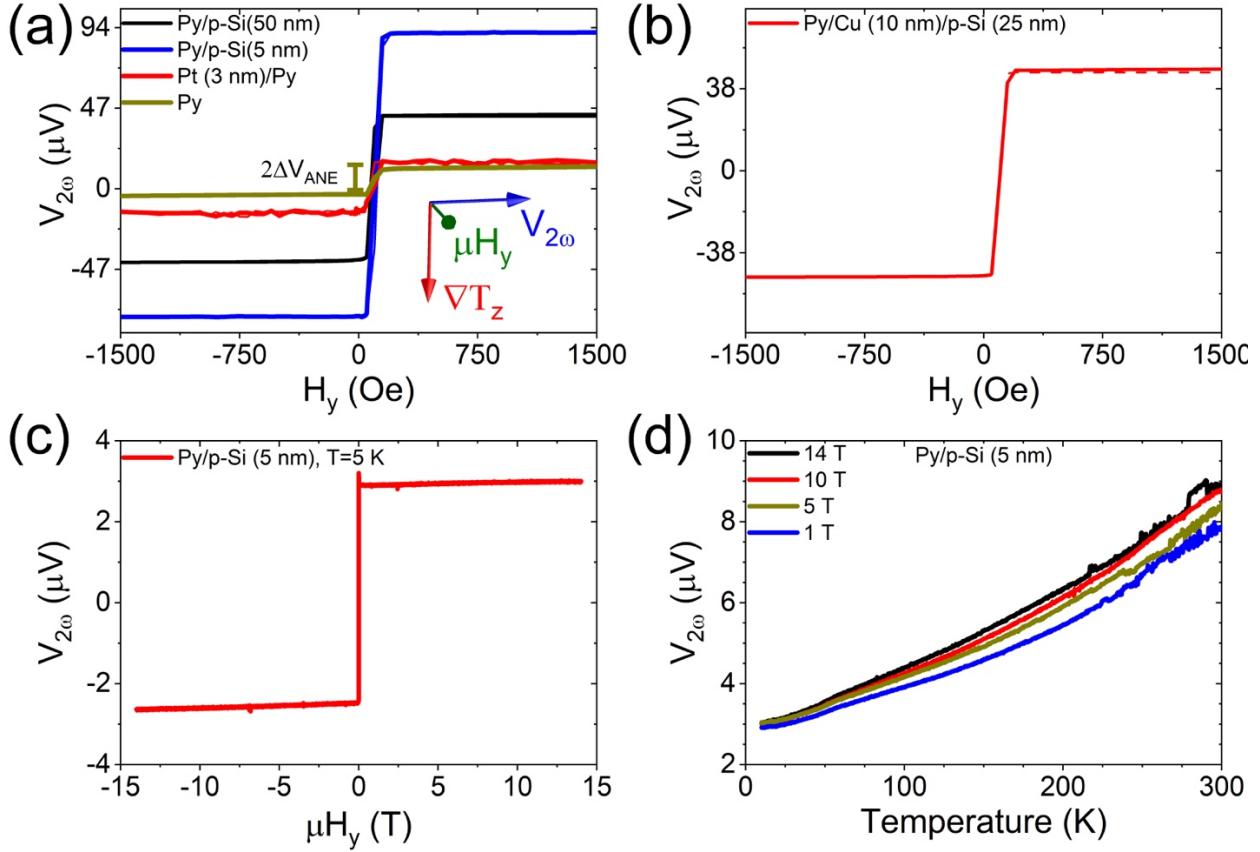

Figure 2. the transverse spin dependent thermoelectric response in (a) Py (25 nm), Pt (3 nm)/Py (25 nm), Py (25 nm)/p-Si (50 nm) and Py (25 nm)/p-Si (5 nm) samples at 300 K, and (b) in Py (25 nm)/Cu (10 nm)/p-Si (25 nm) sample at 300 K. (c) The high magnetic field transverse spin dependent thermoelectric response in Py (25 nm)/p-Si (5 nm) sample at 5 K, and (d) the transverse spin dependent thermoelectric response as a function of temperature in Py (25 nm)/p-Si (5 nm) sample at an applied magnetic field ($\mu H_y$) of 1 T, 5 T, 10 T and 14 T from 300 K to 10 K. The fluctuations in the temperature-dependent measurements are due to instrumental settings.

The large response in the p-Si samples was expected to arise due to topological electronic magnetism of phonons. However, the measured response could have come from the proximity magnetism. We, then, introduced a 10 nm Cu layer in between the Py (25 nm) and Si (25 nm) to remove any proximity driven effect[18,19]. For similar heating power, we measured the $V_{2\omega}$ response to be 95.8 µV in this sample as shown in Figure 2 (b), which is larger than that of Pt sample. The addition of Cu layer should have reduced the transverse thermal response due to shunting effect. However, the larger response in this case might arise from an additional strain gradient and large flexoelectronic doping in 25 nm p-Si layer from the Cu layer. The enhanced thermal response in Si samples cannot be attributed to interfacial engineering of ANE behavior[20] since, in that case, the response should have been similar for both 5 nm and 50 nm p-Si. Additionally, the enhanced ANE response should have disappeared in the sample with Cu interlayer, which did not happen.

Next, we measured the $V_{2\omega}$ response at 5 K for 10 mA of heating current and applied magnetic field from 14 T to -14 T in the 5 nm Si bilayer sample, as shown in Fig. 2 (c). The measurement showed no reduction in response at high magnetic field. Hence, the origin of the response was electronic and not the magnonic spin current [21,22]. This assertion was further supported by the $V_{2\omega}$ response as a function of temperature from 10 K to 300 K with applied magnetic field 1 T, 5 T, 10 T, and 14 T as shown in Fig. 2 (d). The $V_{2\omega}$ response increased as the magnetic field is increased. The increase in the $V_{2\omega}$ response with increasing magnetic field can be attributed to reduction in electron-magnon scattering at higher magnetic fields. The reduction in electron-magnon scattering was expected to increase the spin polarization of Py layer, which increased the flexoelectronic spin current.

We attributed the spin dependent thermoelectric response to topological electronic magnetism of phonons. topological electronic magnetism of phonons, which is described as:

$$\mathbf{M_t} \propto \frac{\partial n}{\partial z} \times f(\mathbf{A}, \mathbf{p}) \tag{1}$$

where $M_t$ is temporal magnetic moment, $\mathbf{P_{FE}} \propto \frac{\partial n}{\partial z}$ is flexoelectronic effect due to strain gradient that give rise to gradient of charge carrier concentration$\left(\frac{\partial n}{\partial z}\right)$ and $\partial_t \mathbf{P} \propto f(\mathbf{A}, \mathbf{p})$ is time evolution of topological phonon polarization that is a function of Berry gauge potential $\mathbf{A}$ and momentum $\mathbf{p}$. However, the p-Si layer in the samples was deposited using RF sputtering. The roughness of the sample was expected to be ~1.2 nm as stated earlier. The charge carrier concentration was inhomogeneous due to thickness variation and strain gradient variation. The equation 1 can be rewritten as:

$$\mathbf{M_t}(x, y, z) \propto \frac{\partial n\ (x,y,z)}{\partial z} \times f(\mathbf{A}, \mathbf{p}) \tag{2}$$

Hence, the topological electronic magnetic moment was expected to be random but having a component perpendicular to the interface based on our previous measurements. The thermoelectric response measurement as a function of magnetic field perpendicular to the interface ($\mu H_z$) for in-plane temperature gradient($\Delta T_y$) should give rise to topological Nernst effect (TNE) behavior. Hence, we undertook measurements in perpendicularly magnetized (PM) configuration[12] as shown in Figure 3 (a) inset. In the measurement devices, the heater and sample had an offset due to lithographic misalignment, as shown in Figure 1 (d). This offset gave rise to a temperature difference along the y-axis. The transverse thermoelectric responses in all the samples were measured in PM configuration as shown in Figure 3.

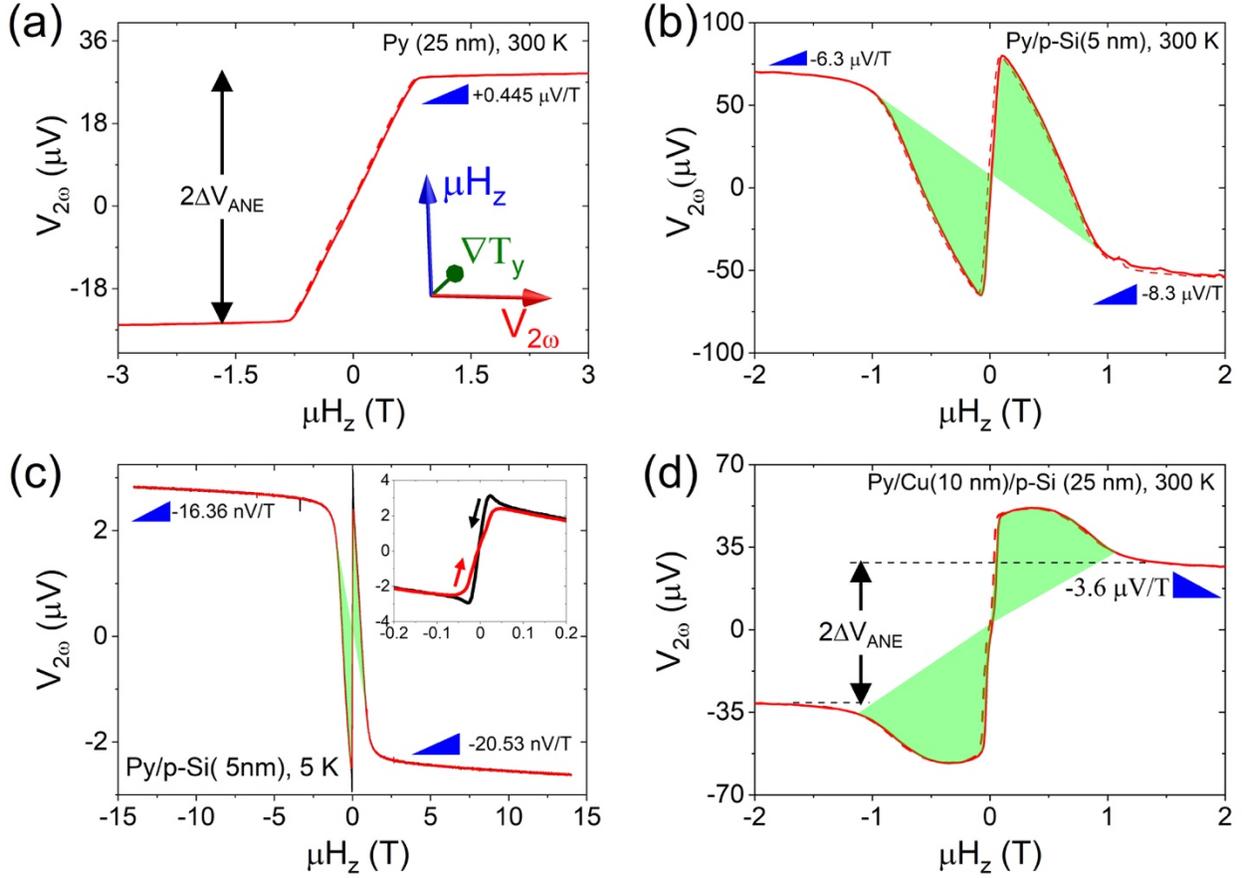

Figure 3. The $V_{2\omega}$ response at 300 K as a function of an applied out of plane magnetic field sweep in (a) Py (25 nm) sample from 3 T to -3 T and (b) in Py (25 nm)/Si ( 5nm) sample from 2 T to -2 T. (c) the $V_{2\omega}$ response at 5 K as a function of an applied out of plane magnetic field sweep in (a) Py (25 nm)/Si (5 nm) sample from 14 T to -14 T, and inset shows the low field behavior. (d) The $V_{2\omega}$ response at 300 K as a function of an applied out of plane magnetic field sweep in Py (25 nm)/Cu (10 nm)/p-Si (25 nm) sample from 2 T to -2 T. The green shaded region represents the possible topological Nernst effect response after subtracting ordinary and anomalous Nernst responses.

The transverse thermoelectric response in Py (25 nm) control sample is measured for applied magnetic field of 3 T to -3 T as shown in Figure 3 (a). The overall response is a combination of anomalous Nernst effect (ANE) and ordinary Nernst effect (ONE) as expected. The sign of both the ANE and ONE is positive. The ANE response in Py was attributed to the intrinsic origin[23]. The ONE response primarily depends on the scattering behavior only. The ONE response in Py sample is attributed to the impurity scattering and ONE response for positive and negative magnetic fields are identical (within measurement error) as shown in Figure 3 (a).

The transverse thermoelectric response in Py (25 nm)/Si (5 nm) sample for applied magnetic field of 2 T to -2 T is shown in Figure 3 (b). The behavior in presence of additional Si layer changed significantly. The overall response can be described as:

$$V_{2\omega} = V_{ONE}^{Py} + V_{ONE}^{Si} + V_{ANE}^{Py} + V_{TNE}^{Si} \qquad (1)$$

The sign of the ONE response is negative as compared to positive in Py control sample. Additionally, the ONE response is asymmetric for positive (-8.3 µV/T) and negative (-6.3 µV/T) magnetic fields as shown in Figure 3 (b). The negative ONE response in Si is attributed to acoustic phonon scattering[24]. The sign of the ANE response was also switched to negative. We attributed the sign reversal in the ANE response to the change in Fermi level due to flexoelectronic charge carrier injection from Py to Si layer as hypothesized, which changed the charge carrier from electrons to holes. We have already demonstrated sign reversal of Hall resistance as well as anomalous Hall resistance in the part 3 of this series, which supported our argument. Our argument was further supported by the observation of sign reversal of anomalous Hall resistance due to gate bias in magnetic topological insulator ($MnBi_2Te_4$)[25]. In the thermoelectric measurement, we also found an additional response (shaded green region) in this sample, which disappears at large magnetic fields as shown in Figure 3 (b). The behavior is similar to topological Hall effect (THE)[26-28] due to topological spin texture since it vanishes at high magnetic field. The thermal equivalent to THE is TNE. Hence the measured response was TNE due to randomly oriented magnetic moments from topological electronic magnetism of phonons as hypothesized.

We, then, measured the transverse thermoelectric response at 5 K for a heating current of 10mA and from 14 T to -14 T as shown in Figure 3 (c). The overall behavior remains same at 5 K. The sign of both the ANE and ONE response is negative. The asymmetric behavior between positive and negative field ONE responses is also similar to the behavior observed at 300 K. More importantly, the TNE like response still exist at low temperature. It meant that the behavior did not arise due to optical phonons and acoustic phonons could be the potential reason for the behavior due to topological Berry gauge potential. The measurement was carried out at large field (14 T) to uncover any additional topological response. However, no additional behavior such as oscillating Nernst effect was observed as shown in Figure 3 (c).

In the Py/p-Si samples, the interfacial flexoelectronic effect was at Py/p-Si interface, which led to charge carrier transfer from Py to p-Si layer and sign reversal of ANE response intrinsic to Py layer. In case of Py/Cu/p-Si sample, Py/Cu interface was not expected to have flexoelectronic polarization. Hence, the ANE response would be positive similar to Py only sample (Figure 3 (a)). Whereas ONE response would be combination of three layers and might still be negative. The measurement in a Py (25 nm)/Cu (10 nm)/ Si (25 nm) is shown in Figure 3 (d). As expected, the sign of ANE is now positive because flexoelectronic charge doping was due to the Cu layer and not from the Py layer. But, the ONE response remains negative except the magnitude was same for both positive and negative fields. More importantly, the TNE response did not disappear even though Py is not in contact with Si. Hence, the TNE response was not from the Py layer or from interfacial alloying.

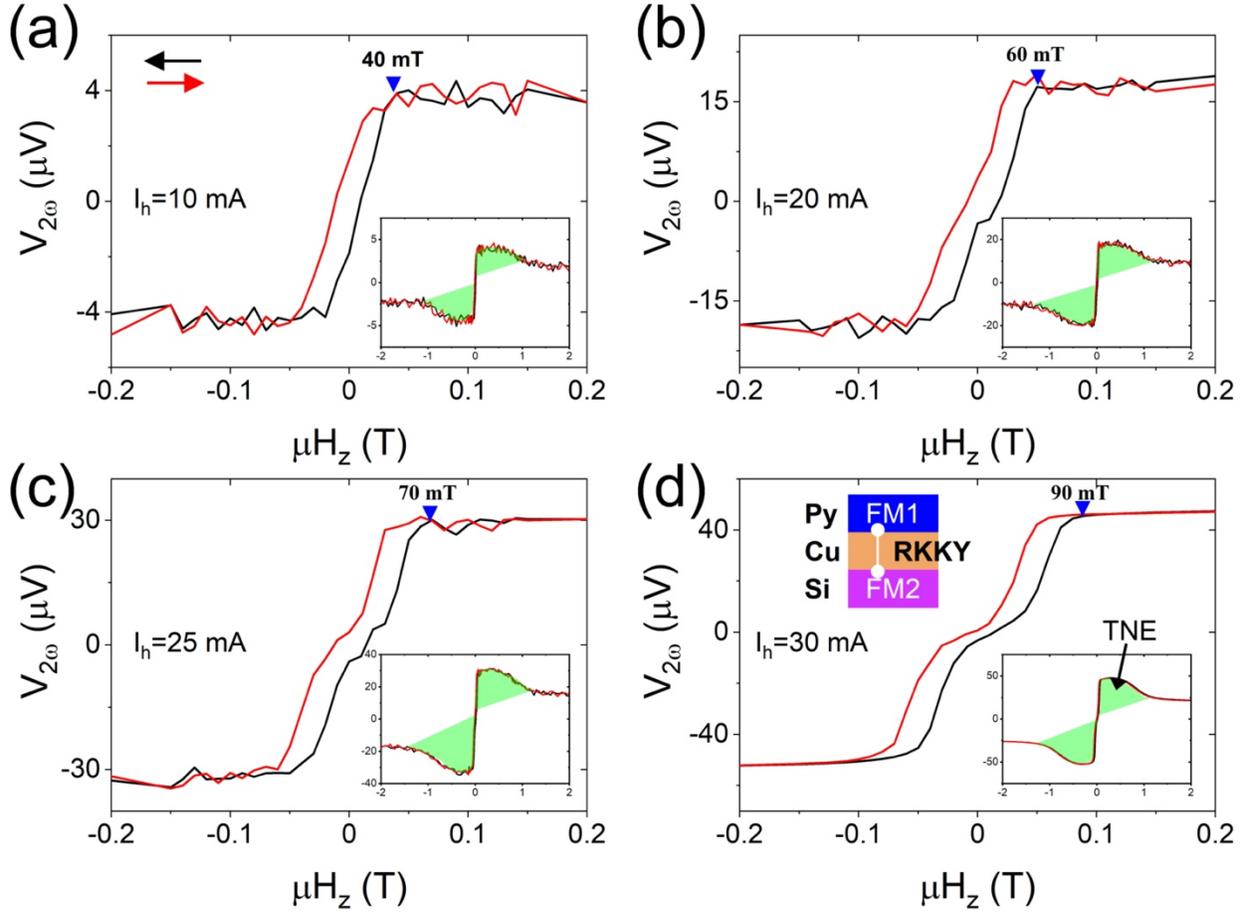

Figure 4. The transverse thermoelectric response in Py (25 nm)/Cu (10 nm)/ Si (25 nm) sample at (a) 10 mA, (b) 20 mA, (c) 25 mA and (d) 30 mA of heating current for an applied magnetic field of 2 T to -2 T. The green shaded area represents possible TNE response.

To further investigate the origin of TNE like behavior, we measured the response at different heating current. The change in heating will also change the thermal expansion stresses, strain gradient and resulting flexoelectronic effect in the Si layer. The transverse thermoelectric response was measured in Py (25 nm)/Cu (10 nm)/ Si (25 nm) sample at 10 mA, 20 mA, 25 mA and 30 mA of heating current and low field behavior from this measurement is shown in Figure 4 (a-d) with overall response shown in the corresponding inset figures. We find that the saturation field increases from 40 mT to 90 mT when the heating current is increased from 10 mA to 30 mA. We attribute this behavior to increase in topological magnetic moment of phonon in Si due to increased flexoelectronic doping. However, the most striking change was observed in switching behavior. The switching behavior shows a transition to an exchange bias mediated behavior as the heating current is increased from 10 mA to 30 mA. The behavior can be attributed to antiferromagnetic RKKY interlayer coupling. However, for RKKY interlayer coupling, two magnetic moments are required. In our sample, the first magnetic moment is Py and second magnetic moment was the magnetic moment of topological phonons in the Si as shown in Figure 4 (d). We did not observe any exchange bias behavior in the in-plane magnetization configuration as shown in Figure 2 (b). This is an unambiguous proof of the randomized magnetic moments of topological phonons having an out of plane net component in the Si layer, which is the underlying cause of TNE response observed in our study. Additionally, we did not observe any such response in control Cu (10 nm)/p-Si (20 nm) sample as shown in Supplementary Figure S5 where very small ONE response (in nV) was observed, which also proved that all the measured responses were of thermal origin.

We, then, made a control device where position of heater and sample was switched as shown in Supplementary Figure S6, which removed any strain gradient from the sample. The transverse thermoelectric response measurement on an unstrained Si (50 nm)/Py (25 nm) sample showed positive sign for both ONE and ANE responses as shown in Supplementary Figure S6. In addition, the TNE like response was completely absent. This control experiment clearly showed the strain gradient is the underlying macroscopic cause of the observed TNE behavior.

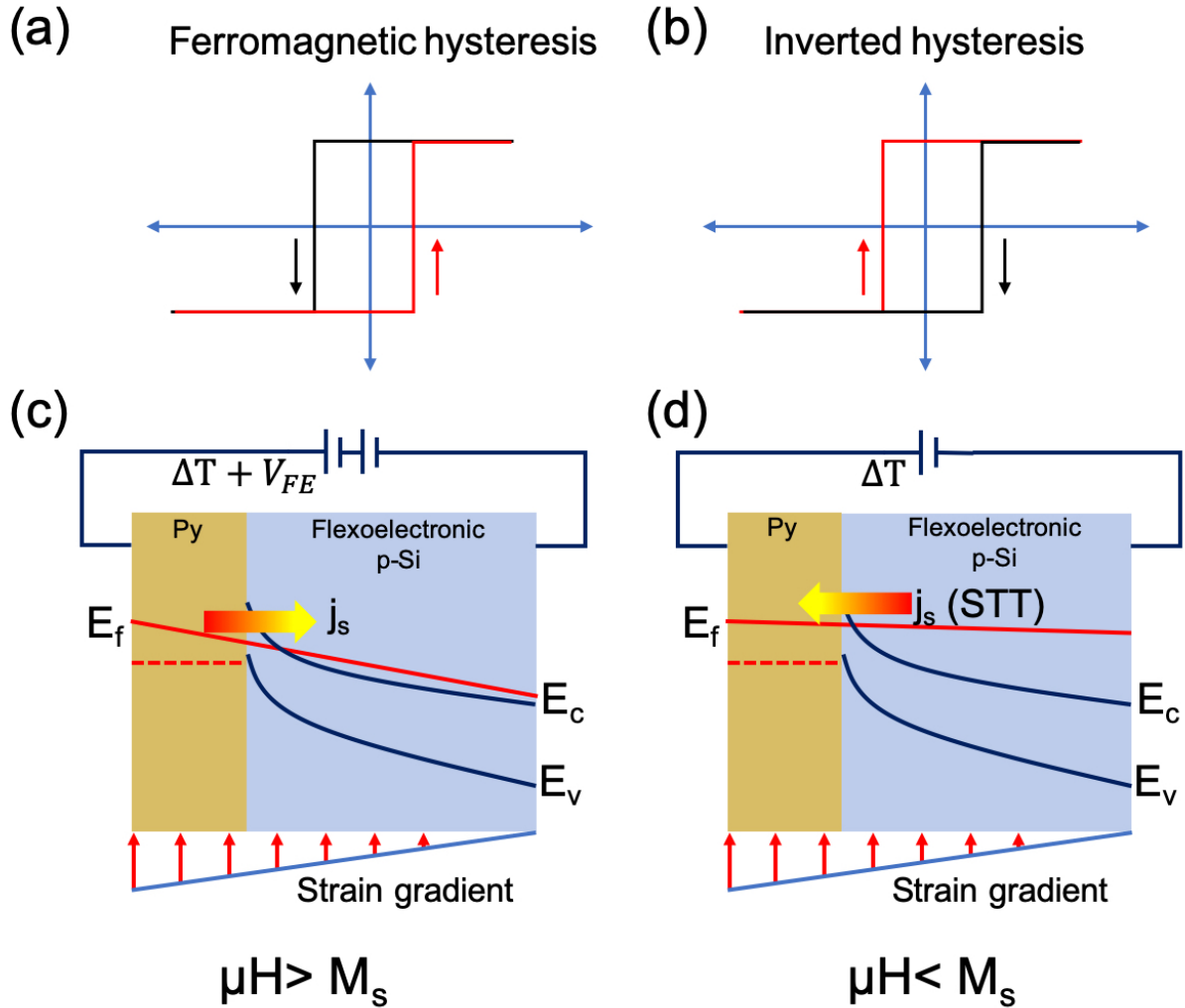

Figure 5. (a) a schematic showing the ferromagnetic switching hysteresis loop, (b) a schematic showing the inverted hysteresis loop, (c) a schematic showing the spin current generated at high magnetic field, and (d) a schematic showing the spin transfer torque generated at low field.

The hysteresis behavior observed in all the p-Si bilayer samples was inverted as shown in Figure 4. The conventional ferromagnetic switching occurs when the applied magnetic field is larger than coercive field as shown in Figure 5 (a). In the inverted hysteretic behavior, the switching occurs even before the magnetic field has switched the direction. The inverted hysteresis is usually attributed to antiferromagnetic interlayer coupling. We have already demonstrated antiferromagnetic RKKY interlayer coupling. However, the flexoelectronic charge transfer was expected to be the primary reason for the inverted hysteresis behavior. At magnetic field larger than saturation magnetic field, the flexoelectronic charge transfer led to a large spin current as well as shown in Figure 5 (c). But, once the applied external magnetic field reduced below the saturation magnetic field, the flexoelectronic spin current was reduced because of randomization of magnetic domains in Py layer. As a consequence, an inverse spin current was generated from the p-Si layer to Py layer as shown in Figure 5 (d), which led to the spin transfer torque (STT) and switching of the Py layer. The flexoelectronic STT was the underlying cause of the inverted hysteresis behavior. More importantly, the flexoelectronic STT can lead to either charge based or thermal based magnetic switching behavior, which can lead to efficient spintronics and spin caloritronics devices.

The experimental results presented in this study demonstrated large spin dependent thermoelectric response. The next step was quantitative description of the spin-to-charge conversion efficiency. The spin-Hall angle of the Pt was 0.068[29] for Py as a ferromagnetic source. We estimated that the SSE response of 5 nm a-Si was 8.5 times that of SSE in Pt sample. The corresponding spin-Hall angle for 5 nm a-Si will be 0.578. This value was much larger than that of most heavy metals[30]. The effective spin mixing conductance in bulk Si ($1.74 \times 10^{19}$ m$^{-2}$ – $5.2 \times 10^{19}$ m$^{-2}$)[31,32] was reported to be similar to that of Pt ($2.1 \times 10^{19}$ m$^{-2}$) [33]. However, there are no measurements for spin injection efficiency for flexoelectronic Si, which was expected to be the underlying cause of large thermoelectric response. As a consequence, the coefficient of spin dependent thermoelectric response as the better parameter for comparison as compared to spin-Hall angle. The largest spin-Seebeck coefficient in this work was estimated to be 1.032 µV/K, which was smaller than YIG/NiO/Pt system reported by Lin et al[34]. Our value was similar to $Fe_3O_4$/Pt system[35]. However, we achieved it using strain gradient and in spite of insignificant SOC in Si.

In conclusion, we reported one of the largest spin dependent thermoelectric response in an inhomogeneously strained metal/p-Si bilayer structure. The spin dependent response is found to be an order of magnitude larger than that in Pt thin films and similar to topological insulators surface states in spite of negligible intrinsic spin-orbit coupling of Si. This large response is attributed to the topological electronic magnetism of phonons, which was uncovered using topological Nernst effect measurement.

**Author contributions**

The manuscript was written through contributions of all authors. All authors have given approval to the final version of the manuscript. ‡RGB and AK have equal contribution to this work.

**Acknowledgement**

The fabrication of experimental devices was done at the Center for Nanoscale Science and Engineering at UC Riverside. The electron microscopy sample preparation and imaging were done at the Central Facility for Advanced Microscopy and Microanalysis at UC Riverside. SK acknowledges research gift from Dr. S Kumar.

**Conflict of interest**

Authors declare no conflict of interest.

**References**

[1]     Y. Ren, C. Xiao, D. Saparov, and Q. Niu, arXiv preprint arXiv:2103.05786 (2021).

[2]     P. C. Lou, A. Katailiha, R. G. Bhardwaj, W. P. Beyermann, D. M. Juraschek, and S. Kumar, Nano Letters **21**, 2939 (2021).

[3]     P. C. Lou, A. Katailiha, R. G. Bhardwaj, T. Bhowmick, W. P. Beyermann, R. K. Lake, and S. Kumar, Phys. Rev. B **101**, 094435 (2020).

[4]     P. C. Lou, L. de Sousa Oliveira, C. Tang, A. Greaney, and S. Kumar, Solid State Communications **283**, 37 (2018).

[5]     L. Wang, S. Liu, X. Feng, C. Zhang, L. Zhu, J. Zhai, Y. Qin, and Z. L. Wang, Nature Nanotechnology **15**, 661 (2020).

[6]     K. Uchida, S. Takahashi, K. Harii, J. Ieda, W. Koshibae, K. Ando, S. Maekawa, and E. Saitoh, Nature **455**, 778 (2008).

[7]     K. Uchida *et al.*, Nat Mater **9**, 894 (2010).

[8]     L. W. Martin and A. M. Rappe, Nature Reviews Materials **2**, 16087 (2017).

[9]     R. G. Bhardwaj, P. C. Lou, and S. Kumar, Appl. Phys. Lett. **112**, 042404 (2018).

[10]    R. G. Bhardwaj, P. C. Lou, and S. Kumar, physica status solidi (RRL) – Rapid Research Letters **12**, 1800064 (2018).

[11]    M. Schmid, S. Srichandan, D. Meier, T. Kuschel, J. M. Schmalhorst, M. Vogel, G. Reiss, C. Strunk, and C. H. Back, Phys. Rev. Lett. **111**, 187201 (2013).

19
**References**

[1]     Y. Ren, C. Xiao, D. Saparov, and Q. Niu, arXiv preprint arXiv:2103.05786 (2021).

[2]     P. C. Lou, A. Katailiha, R. G. Bhardwaj, W. P. Beyermann, D. M. Juraschek, and S. Kumar, Nano Letters **21**, 2939 (2021).

[3]     P. C. Lou, A. Katailiha, R. G. Bhardwaj, T. Bhowmick, W. P. Beyermann, R. K. Lake, and S. Kumar, Phys. Rev. B **101**, 094435 (2020).

[4]     P. C. Lou, L. de Sousa Oliveira, C. Tang, A. Greaney, and S. Kumar, Solid State Communications **283**, 37 (2018).

[5]     L. Wang, S. Liu, X. Feng, C. Zhang, L. Zhu, J. Zhai, Y. Qin, and Z. L. Wang, Nature Nanotechnology **15**, 661 (2020).

[6]     K. Uchida, S. Takahashi, K. Harii, J. Ieda, W. Koshibae, K. Ando, S. Maekawa, and E. Saitoh, Nature **455**, 778 (2008).

[7]     K. Uchida *et al.*, Nat Mater **9**, 894 (2010).

[8]     L. W. Martin and A. M. Rappe, Nature Reviews Materials **2**, 16087 (2017).

[9]     R. G. Bhardwaj, P. C. Lou, and S. Kumar, Appl. Phys. Lett. **112**, 042404 (2018).

[10]    R. G. Bhardwaj, P. C. Lou, and S. Kumar, physica status solidi (RRL) – Rapid Research Letters **12**, 1800064 (2018).

[11]    M. Schmid, S. Srichandan, D. Meier, T. Kuschel, J. M. Schmalhorst, M. Vogel, G. Reiss, C. Strunk, and C. H. Back, Phys. Rev. Lett. **111**, 187201 (2013).

[12]     T. Kikkawa, K. Uchida, S. Daimon, Y. Shiomi, H. Adachi, and Z. Qiu, Phys. Rev. B **88**, 214403 (2013).

[13]     D. G. Cahill, Rev. Sci. Instrum. **61**, 802 (1990).

[14]     T. C. Chuang, P. L. Su, P. H. Wu, and S. Y. Huang, Phys. Rev. B **96**, 174406 (2017).

[15]     J. Holanda, O. Alves Santos, R. O. Cunha, J. B. S. Mendes, R. L. Rodríguez-Suárez, A. Azevedo, and S. M. Rezende, Phys. Rev. B **95**, 214421 (2017).

[16]     S. H. Wang, L. K. Zou, J. W. Cai, B. G. Shen, and J. R. Sun, Phys. Rev. B **88**, 214304 (2013).

[17]     W. Wang *et al.*, Nature Nanotechnology **14**, 819 (2019).

[18]     D. Tian, Y. Li, D. Qu, X. Jin, and C. L. Chien, Appl. Phys. Lett. **106**, 212407 (2015).

[19]     T. Kikkawa, K. Uchida, Y. Shiomi, Z. Qiu, D. Hou, D. Tian, H. Nakayama, X. F. Jin, and E. Saitoh, Phys. Rev. Lett. **110**, 067207 (2013).

[20]     K.-i. Uchida, T. Kikkawa, T. Seki, T. Oyake, J. Shiomi, Z. Qiu, K. Takanashi, and E. Saitoh, Phys. Rev. B **92**, 094414 (2015).

[21]     H. Jin, S. R. Boona, Z. Yang, R. C. Myers, and J. P. Heremans, Phys. Rev. B **92**, 054436 (2015).

[22]     T. Kikkawa, K.-i. Uchida, S. Daimon, Z. Qiu, Y. Shiomi, and E. Saitoh, Phys. Rev. B **92**, 064413 (2015).

[23]     N. Nagaosa, J. Sinova, S. Onoda, A. H. MacDonald, and N. P. Ong, Reviews of Modern Physics **82**, 1539 (2010).

[24]     J. E. Parrott, Proceedings of the Physical Society **71**, 82 (1958).

[25] S. Zhang *et al.*, Nano Letters **20**, 709 (2020).

[26] J. Matsuno, N. Ogawa, K. Yasuda, F. Kagawa, W. Koshibae, N. Nagaosa, Y. Tokura, and M. Kawasaki, Science Advances **2** (2016).

[27] L. Vistoli *et al.*, Nature Physics (2018).

[28] P. Bruno, V. K. Dugaev, and M. Taillefumier, Physical Review Letters **93**, 096806 (2004).

[29] Y. Wang, P. Deorani, X. Qiu, J. H. Kwon, and H. Yang, Appl. Phys. Lett. **105**, 152412 (2014).

[30] J. Sinova, S. O. Valenzuela, J. Wunderlich, C. H. Back, and T. Jungwirth, Rev. Mod. Phys. **87**, 1213 (2015).

[31] Y.-C. Weng, C. T. Liang, and J. G. Lin, Appl. Phys. Lett. **115**, 232101 (2019).

[32] E. Shikoh, K. Ando, K. Kubo, E. Saitoh, T. Shinjo, and M. Shiraishi, Phys. Rev. Lett. **110**, 127201 (2013).

[33] O. Mosendz, J. E. Pearson, F. Y. Fradin, G. E. W. Bauer, S. D. Bader, and A. Hoffmann, Phys. Rev. Lett. **104**, 046601 (2010).

[34] W. Lin, K. Chen, S. Zhang, and C. L. Chien, Phys. Rev. Lett. **116**, 186601 (2016).

[35] R. Ramos *et al.*, Appl. Phys. Lett. **102**, 072413 (2013).

**Supplementary information: Topological phonons in an inhomogeneously strained silicon-4: Large spin dependent thermoelectric response and thermal spin transfer torque due to topological electronic magnetism of phonons**

Ravindra G Bhardwaj[1‡], Anand Katailiha[1‡], Paul C. Lou[1], Ward P. Beyermann[2] and Sandeep Kumar[1, *]

[1] Department of Mechanical Engineering, University of California, Riverside, CA 92521, USA

[2] Department of Physics and Astronomy, University of California, Riverside, CA 92521, USA

**Supplementary Section A- Device fabrication process and experimental measurement**

We take a prime Si wafer and deposit 350 nm of thermal silicon oxide using chemical vapor deposition (CVD). Using lift-off photolithography, we then deposit the sample to be studied using the RF sputtering. The sputtering deposition will have substrate conformal thin film coating. Hence, it will have the same interfacial and surface roughness as the underlying layers. The p-Si target used to deposit amorphous-Si layer is Boron-doped with resistivity of 0.005-0.01 Ω-cm. The second lift-off photolithography is carried out to deposit 50 nm MgO to electrically isolate the sample from the heater. The third lift-off photolithography is then used to deposit heater composed of Ti (10 nm)/Pt (100 nm) using e-beam evaporation. A representative image is shown in Figure 1 (d). The temperature gradient ($\Delta T_z$) was generated across the thickness of the thin film sample by passing an electric current (I) through the Pt heater, as shown in Figure 1 (d).

The residual stresses from Pt (heater) and MgO (insulator) layers are proposed to be the underlying cause of strain gradient mediated Rashba SOC. The existence of residual stresses can be seen in Supplementary Figure S1, where the Pt heater was delaminated due to the residual stresses.

For the second set of devices (unstrained) with switched sample and heater positions, we first deposited Ti (10 nm) / Pt (100 nm) on a Si wafer with predisposition of thermal silicon oxide (650 nm) using CVD. We then sputter 50 nm MgO using RF sputtering for electrical isolation. We fabricated two set of devices having the p-Si (50 nm)/Py (25 nm) bilayer sample and Py (25 nm) sample on top of the MgO.

All the measurements were carried out at 30 mA of heating current except for strained Py/Pt device where measurement was carried out at 25 mA. However, the measured response at 25 mA was multiplied with a factor of 1.44 to get the equivalent response to 30 mA. This was done to protect the device and to have a complete measurement. The Py/Pt devices were found to be fragile for larger current, which could be due to larger residual stresses. The $V_{2\omega}$ response was measured using lock-in technique.

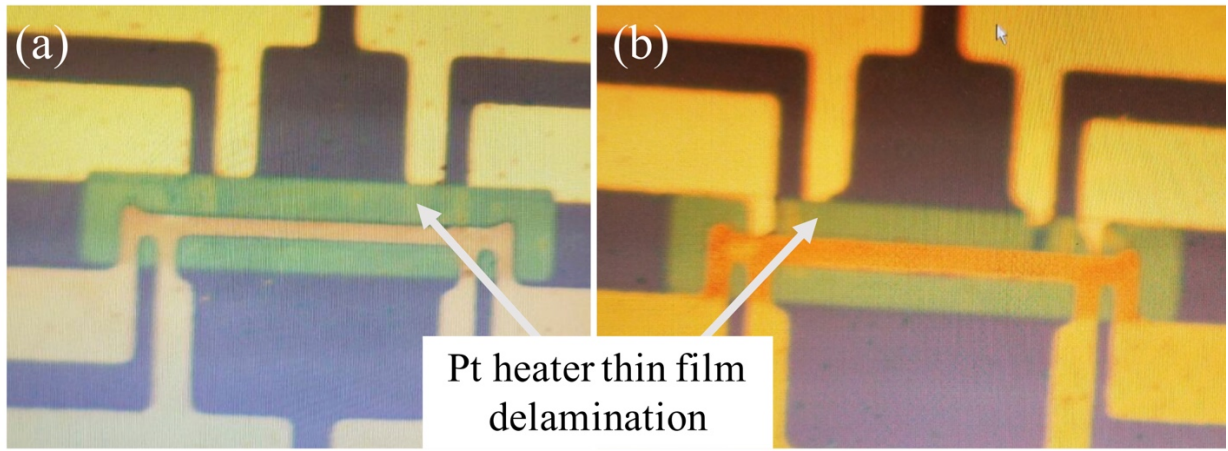

Supplementary Figure S1. (a)-(b)The optical images showing Pt thin film heater layer peeled off (delaminated) due to residual stresses for two devices. The residual stress in Pt heater and MgO layer is proposed to be the cause of strain and strain gradient in the underlying sample.

**Supplementary Section B- Sample characterization**

TEM sample preparation- TEM lamellae were prepared from the layered sample following established procedures with a Dual Beam scanning electron microscope and FIB instrument using Ga ion source (Quanta 200i 3D, Thermo-Fisher Scientific). First, a strap of 5 µm thick protective Carbon layer was deposited over a region of interest using the ion beam. Subsequently approximately 80 nm thin lamella of was cut and polished at 30 kV and attached to a TEM grid using in-situ Omniprobe manipulator. To reduce surface amorphization and Gallium implantation final milling at 5 kV and 0.5 nA was used to thin the sample further.

S/TEM imaging and analysis- TEM and STEM imaging was performed at 300 kV accelerating voltage in a Thermo-Fisher Scientific Titan Themis 300 instrument, fitted with X-FEG electron source, 3 lens condenser system and S-Twin objective lens. High-resolution TEM images were recorded at resolution of 2048x2048 pixels with a FEI CETA-16M CMOS digital camera with beam convergence semi-angle of about 0.08 mrad. STEM images were recorded with Fischione Instruments Inc. Model 3000 High Angle Annular Dark Field (HAADF) Detector with probe current of 150 pA, frame size of 2048x2048, dwell time of 15 µsec/pixel, and camera length of 245 mm. Energy dispersive X-ray Spectroscopy (EDS) analyzes and elemental mapping were obtained in the STEM at 300 kV, utilizing Thermo-Fisher Scientific SuperX system equipped with 4x30mm$^2$ window-less SDD detectors symmetrically surrounding the specimen with a total collection angle of 0.68 srad, by scanning the thin foil specimens. Elemental mapping was performed with an electron beam probe current of 550 pA at 1024 x1024 frame resolution.

Atomic force microscope (AFM) characterization of surface roughness- The surface roughness of the bilayer sample directly reflects the underlying interfacial roughness. The interfacial roughness cannot be more than the surface roughness since the sputter coating is conformal. The AFM measurements are carried out on samples having 50 nm p-Si and 5 nm p-Si layers as shown in Supplementary Figure S2.

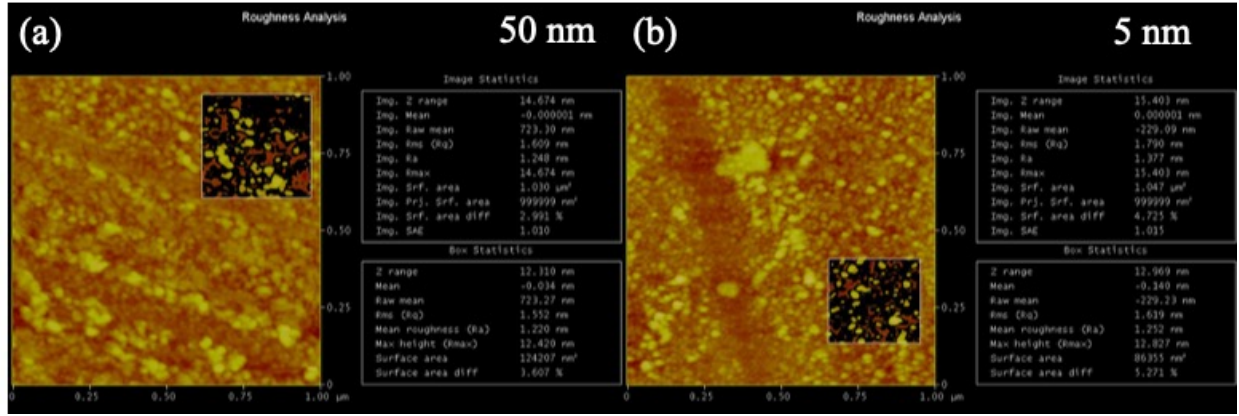

Supplementary Figure S2. The AFM measurements at the surface of Py layer in (a) 50 nm and (b) 5 nm p-Si bilayer samples. The mean roughness of both samples is ~1.2 nm.

**Supplementary Section C- The temperature estimation and spin-Seebeck coefficient calculation**

The increase in temperature across a current carrying heater (Pt) can be estimated using 3ω method, which is given by following equation:

$$\Delta T = \frac{4V_{3\omega}}{R' I_{rms}} \qquad (S1)$$

where, $V_{3\omega}$ is the third harmonic response, $R'$ is the resistance as a function of temperature, and $I_{rms}$ is the heating current. The $3\omega$ measurement was carried out at 20 mA and resulting temperature rise was estimated to be 18.817 K and 18.18 K for 50 nm p-Si and 5 nm p-Si devices, respectively. The $V_{3\omega}$ responses were 6.58 mV and 5.0718 mV for 50 nm p-Si and 5 nm p-Si devices, respectively. Similarly, the values of $R'$ were estimated to be 0.07 $\Omega/K$ and 0.055 $\Omega/K$ for Pt thin film in 50 nm p-Si and 5 nm p-Si devices, respectively. Resistivity of the Pt heater is approximately 2.36×10$^{-7}$ $\Omega m$. The corresponding temperature rise for 30 mA of heating current would be 42.34 K and 41.50 K, respectively. Using this temperature information, we did finite element simulation using COMSOL and found the temperature difference across the Py film for 50 nm p-Si and 5 nm p-Si devices to be 11.924 mK and 11.388 mK, respectively. The values of thermal conductivities used in COMSOL simulations are 20 W/mK, 30 W/mK and 15 W/mK for Py, 50 nm p-Si and 5 nm p-Si, respectively. The vertical temperature distribution for 5 nm p-Si and 50nm p-Si along with temperature data at the interfaces is shown in the Supplementary Figure S3 (a,b).

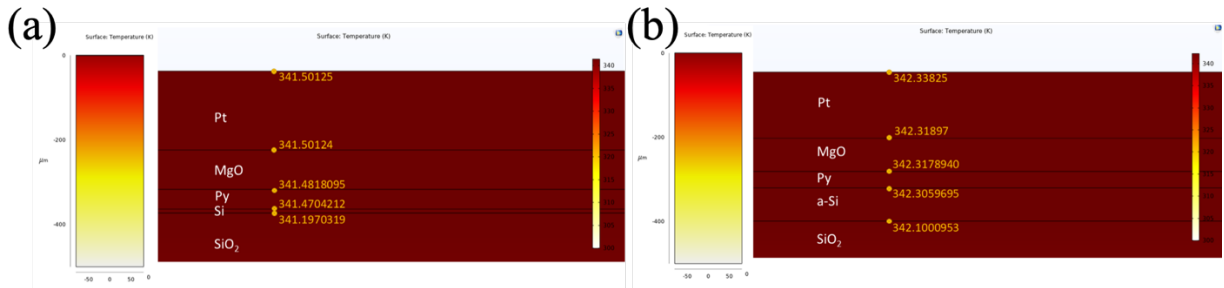

Supplementary Figure S3. The vertical temperature distribution in (a) 5 nm p-Si SSE device and (b) 50 nm p-Si SSE device using COMSOL finite element model.

To demonstrate the effect of magnetic field on heater temperature, we measured the angle dependent $V_{3\omega}$ response for an applied magnetic field of 2 T in yx-plane for Py/p-Si (5 nm) sample. We did not observe significant drift in the heater temperature at 20 mA of heater current as shown in Supplementary Figure S4.

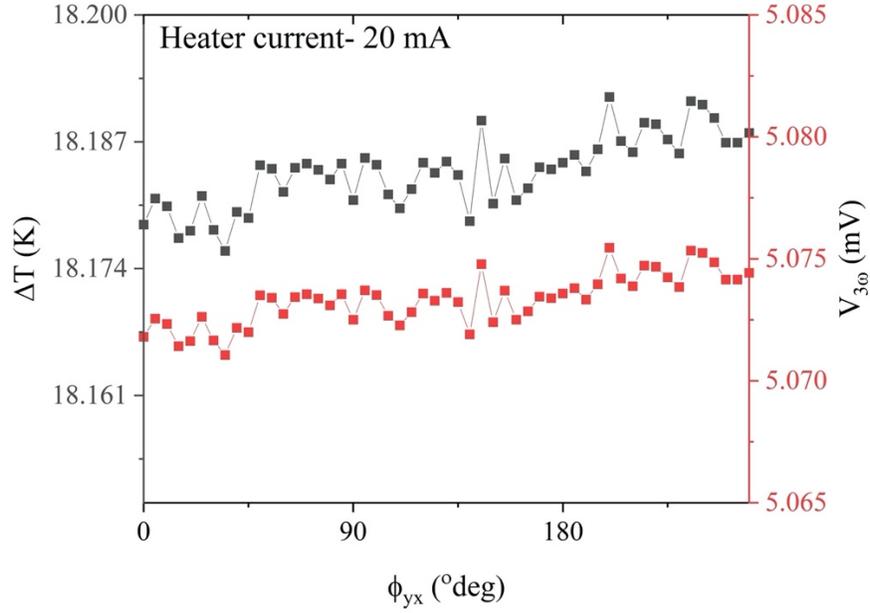

Supplementary Figure S4. The estimated increase in heater temperature and the $V_{3\omega}$ response at an applied magnetic field of 2 T.

The spin Seebeck coefficient are estimated using following equation:

$$S_{LSSE} = \frac{V_{SSE}\, t}{L\Delta T} \qquad (S2)$$

where, $2V_{SSE} = V_{2\omega}$, t = 25 nm, L = 160 μm and $\Delta T$ is the temperature difference across Py film. For 5nm p-Si and 50 nm p-Si the $V_{2\omega}$ is 150.45 μV and 70 μV respectively. So, the $S_{LSSE}$ for 5 nm p-Si and 50nm p-Si is 1.032±0.1 μV/K and 0.458±0.05 μV/K, respectively.

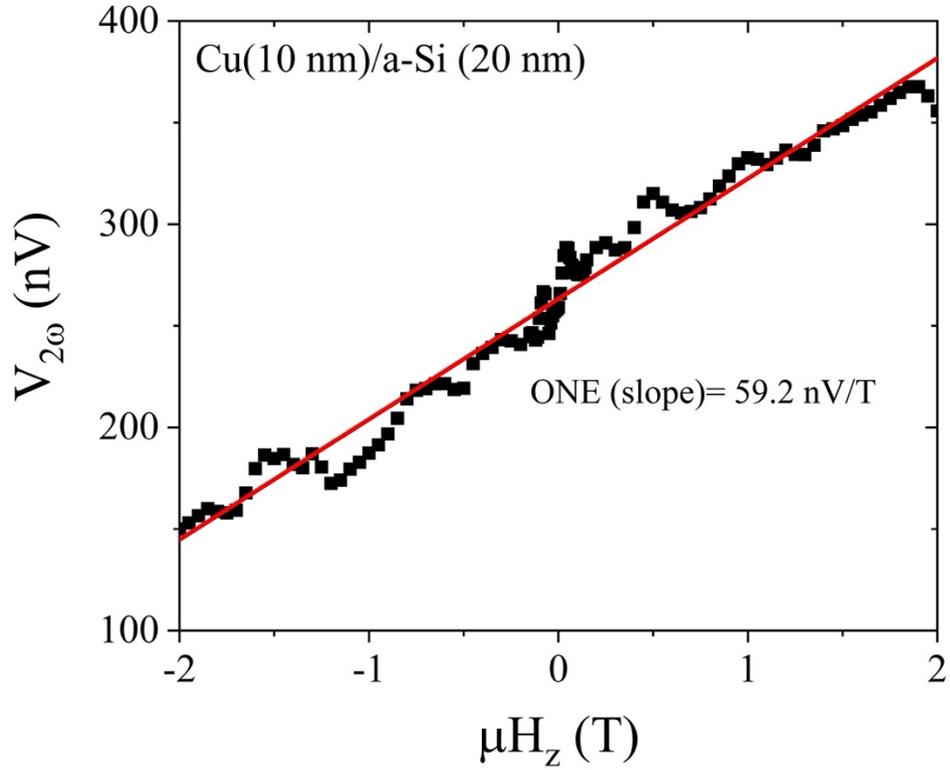

Supplementary Figure S5. the response in PM configuration in Cu (10 nm)/p-Si (20 nm) sample showing ONE behavior.

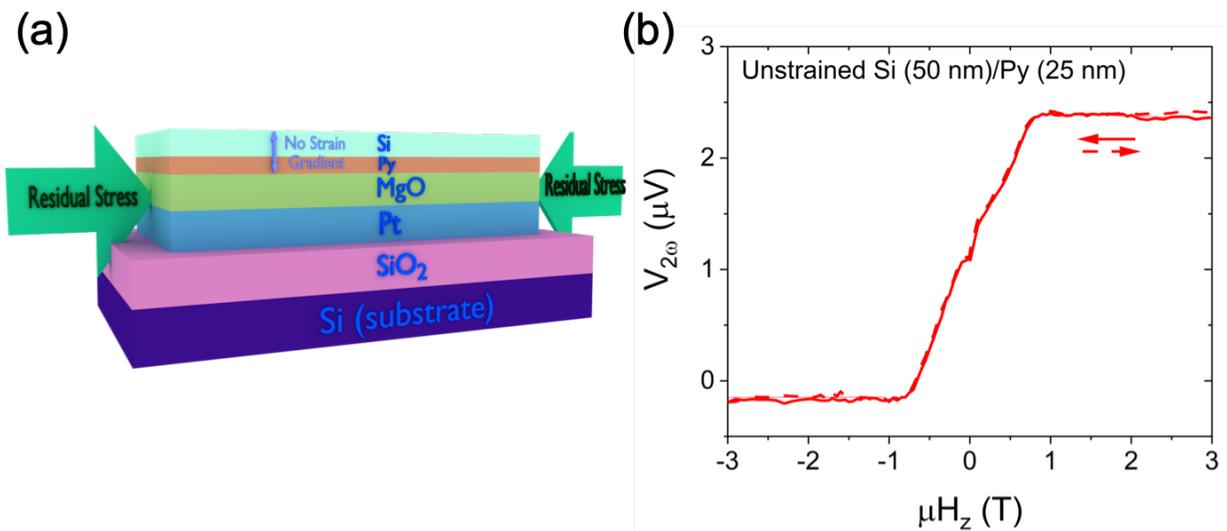

Supplementary Figure S6. (a) A schematic showing the device structure of device with unstrained sample. The position of the sample and heater are switching. Which allows

sample (Si layer) to be unconstrained. (b) the transverse thermoelectric response in Si (50nm)/Py (25 nm) unstrained sample showing positive sign for both ONE and ANE response and the absence of TNE like response.